\documentclass[sigconf,sigplan,screen]{acmart}
\AtBeginDocument{%
  \providecommand\BibTeX{{%
    \normalfont B\kern-0.5em{\scshape i\kern-0.25em b}\kern-0.8em\TeX}}}

\usepackage{subcaption}
\usepackage[ruled]{algorithm2e}
\usepackage{listings}
\usepackage{xcolor}
\usepackage{microtype}
\usepackage{balance}

\definecolor{codegreen}{rgb}{0,0.6,0}
\definecolor{codegray}{rgb}{0.5,0.5,0.5}
\definecolor{codepurple}{rgb}{0.58,0,0.82}
\definecolor{backcolour}{rgb}{0.95,0.95,0.92}
\definecolor{g-blue}{RGB}{66,133,244}
\definecolor{g-green}{RGB}{15, 157, 88}
\definecolor{g-red}{RGB}{219, 68, 55}
\definecolor{g-yellow}{RGB}{244, 180, 0}
\definecolor{f-grey}{RGB}{137, 143, 156}
\definecolor{a-grey}{RGB}{85, 85, 85}
\definecolor{back-grey}{gray}{0.999}
\definecolor{GrayCodeBlock}{RGB}{241,241,241}
\definecolor{BlackText}{RGB}{110,107,94}
\definecolor{RedTypename}{RGB}{39,174,96}
\definecolor{GreenString}{RGB}{96,172,57}
\definecolor{PurpleKeyword}{RGB}{231, 76, 60}
\definecolor{GrayComment}{RGB}{170,170,170}
\definecolor{GoldDocumentation}{RGB}{180,165,45}
\definecolor{BlueKeys}{RGB}{52,152,219}
\definecolor{RedKeys}{RGB}{231, 76, 60}
\definecolor{delim}{RGB}{20,105,176}
\definecolor{numb}{RGB}{255 ,153 ,0}
\definecolor{string}{rgb}{0.64,0.08,0.08}

\lstdefinelanguage{egg}
{   
    breakatwhitespace=false,         
    breaklines=true,                 
    captionpos=b,                    
    keepspaces=true,                 
    numbers=left,                    
    numbersep=5pt,                  
    showspaces=false,                
    showstringspaces=false,
    showtabs=false,         
    frameround=tttt,
    tabsize=2,
    columns=fullflexible,
    postbreak=\raisebox{0ex}[0ex][0ex]{\ensuremath{\color{a-grey}\hookrightarrow\space}},
    columns=fullflexible,
    basicstyle=\ttfamily\color{BlackText},
    keywords={
        unsafe,async,await,move,
        use,pub,crate,super,self,mod,
        struct,enum,fn,const,static,let,mut,ref,type,impl,dyn,trait,where,as,
        break,continue,if,else,while,for,loop,match,return,yield,in,def,function,::, new,
    },
    keywordstyle=\color{PurpleKeyword},
    keywords=[2]{Self, Pattern, Id, Vec, Rewrite, bool,u8,u16,u32,u64,u128,i8,i16,i32,i64,i128,char,str,
        Self,Option,Some,None,Result,Ok,Err,String,Box,Vec,Rc,Arc,Cell,RefCell,HashMap,BTreeMap,
        macro_rules,Pattern, RecExpr, Duration, Runner,  true, false, Math},
    keywordstyle=[2]\color{RedTypename},
    keywords=[3]{idef,not, and, add, merge, check\_equivalence, is\_saturated\_or\_timeout, ematch, search\_eclass, push, to\_string, Other, format!, info!, is\_none, enumerate, iter, into\_iter, collect, rebuild, run\_one, initial\_egraph, find, last, unwrap, default, parse, clone, with_iter_limit, with_node_limit, with_expr, with_time_limit, run, rules, extract_best},
    keywordstyle=[3]\color{BlueKeys},
    comment=[l][\color{GrayComment}\slshape]{//},
    morecomment=[s][\color{GrayComment}\slshape]{/*}{*/},
    morecomment=[l][\color{GoldDocumentation}\slshape]{///},
    morecomment=[s][\color{GoldDocumentation}\slshape]{/*!}{*/},
    morecomment=[l][\color{GoldDocumentation}\slshape]{//!},
    morecomment=[s][\color{RedTypename}]{\#![}{]},
    morecomment=[s][\color{RedTypename}]{\#[}{]},
    stringstyle=\color{GreenString},
    string=[b]",
}

\setcopyright{none}
\acmPrice{15.00}
\acmDOI{10.1145/3497776.3517781}
\acmYear{2022}
\copyrightyear{2022}
\acmSubmissionID{cc22main-p136-p}
\acmISBN{978-1-4503-9183-2/22/04}
\acmConference[CC '22]{Proceedings of the 31st ACM SIGPLAN International Conference on Compiler Construction}{April 02--03, 2022}{Seoul, South Korea}
\acmBooktitle{Proceedings of the 31st ACM SIGPLAN International Conference on Compiler Construction (CC '22), April 02--03, 2022, Seoul, South Korea}

\begin{document}

\title{Caviar: An E-Graph Based TRS for Automatic Code Optimization}

\author{Smail Kourta}
\authornote{Both authors contributed equally to the paper}
\affiliation{
  \institution{New York University Abu Dhabi}
  \country{United Arab Emirates}
}
\affiliation{
  \institution{École nationale supérieure d'informatique}
  \country{Algeria}
}
\email{gs_kourta@esi.dz}

\author{Adel Abderahmane Namani}
\authornotemark[1]  
\affiliation{
  \institution{New York University Abu Dhabi}
  \country{United Arab Emirates}
}
\affiliation{
  \institution{École nationale supérieure d'informatique}
  \country{Algeria}
}
\email{ga_namani@esi.dz}

\author{Fatima Benbouzid-Si Tayeb}      
\affiliation{
  \institution{École nationale supérieure d'informatique}
  \country{Algeria}
}
\email{f_sitayeb@esi.dz}

\author{Kim Hazelwood}      
\affiliation{
  \institution{Meta AI}
  \country{United States of America}
}
\email{kimhazelwood@fb.com}

\author{Chris Cummins}      
\affiliation{
  \institution{Meta AI}
  \country{United States of America}
}
\email{cummins@fb.com}

\author{Hugh Leather} 
\affiliation{
  \institution{Meta AI}
  \country{United States of America}
}
\email{hleather@fb.com}

\author{Riyadh Baghdadi}  

\affiliation{
  \institution{New York University Abu Dhabi}
  \country{United Arab Emirates}
}
\email{baghdadi@nyu.edu}

\renewcommand{\shortauthors}{Kourta and Namani, et al.}

\begin{abstract}
Term Rewriting Systems (TRSs) are used in compilers to simplify and prove expressions.
State-of-the-art TRSs in compilers use a greedy algorithm that applies a set of rewriting rules in a predefined order (where some of the rules are not axiomatic). This leads to a loss of the ability to simplify certain expressions.
E-graphs and equality saturation sidestep this issue by representing the different equivalent expressions in a compact manner from which the optimal expression can be extracted. While an e-graph-based TRS can be more powerful than a TRS that uses a greedy algorithm, it is slower because expressions may have a large or sometimes infinite number of equivalent expressions. Accelerating e-graph construction is crucial for making the use of e-graphs practical in compilers. In this paper, we present Caviar, an e-graph-based TRS for proving expressions within compilers. 
The main advantage of Caviar is its speed. It can prove expressions much faster than base e-graph TRSs.
It relies on three techniques: 1) a technique that stops e-graphs from growing when the goal is reached, called Iteration Level Check; 2) a mechanism that balances exploration and exploitation in the equality saturation algorithm, called Pulsing Caviar; 3) a technique to stop e-graph construction before reaching saturation when a non-provable pattern is detected, called Non-Provable Patterns Detection (NPPD). We evaluate caviar on Halide, an optimizing compiler that relies on a greedy-algorithm-based TRS to simplify and prove its expressions. The proposed techniques allow Caviar to accelerate e-graph expansion for the task of proving expressions. They also allow Caviar to prove expressions that Halide’s TRS cannot prove while being only 0.68x slower. Caviar is publicly available at: \url{https://github.com/caviar-trs/caviar}.
\end{abstract}




\maketitle

\section{Introduction}
Equality graphs (e-graphs) \cite{nelson_fast_1980} are a particular kind of graphs that store a set of terms and the equivalence relation over them.  They were originally developed to efficiently represent congruence relations in automated theorem provers (ATPs). Over the past decade, several projects have re-purposed e-graphs to implement state-of-the-art compiler optimizations and program synthesizers using a technique known as equality saturation. This technique applies all rewrite rules simultaneously until no additional information can be added to the e-graph \cite{tate_equality_2009}, at which point the e-graph is said to be saturated.

E-graphs provide more power than traditional methods for term rewriting, they represent the different equivalent expressions in a compact manner from which the optimal expression can be extracted. But their expansion phase (graph construction) is time-consuming due to the application of all rewrite rules on each iteration of the algorithm. This makes the expansion of the e-graph impractical for most applications. The problem is worse when the size of the ruleset (number of rewriting rules) is large. In some cases, if the ruleset is not chosen carefully, the expansion phase of the e-graph can run indefinitely due to the expression having an infinite number of forms.

In this work, we present an e-graph based TRS (Term Rewriting System) named \emph{Caviar}.
Caviar is designed to prove and simplify expressions using only axiomatic rules (rules that cannot be derived from other rules), thus reducing the size of the ruleset making it easier to maintain and more powerful.
Proving an expression, in this context, means proving whether it evaluates to true or false.
Although Caviar can be used to either simplify or prove expressions, in this paper, we will focus only on its use case to prove expressions.
The three expressions below are examples of expressions that Caviar can be used to prove.\footnote{We chose short expressions as examples, in general, the expressions that compilers need to prove can be much longer.}. These expressions were generated by optimization passes in the Halide compiler~\cite{ragan-kelley_halide_2013}.

\[ (((v_0 + -1) / 2) <= ((((((v_0 + 1) / 2) - v_1) / 2) * 2) + v_1)) \]

\[ (max(((v_0 + -1) / 2), (((v_0 + 1 ) \% 2 ) * 2))) <= ((v0 + 1)/2)) \]

\[ ((((v_0 - v_1) / 8) + 32) <= max((((v_0 - v_1) + 257) / 8), 0 )) \]\\

An expression, similar to the above, is usually generated by a compiler pass that needs to check whether the expression evaluates to true or false. Based on the answer, an optimization might be applied. For example, in order to decide whether an array could be stored in the shared memory of a GPU (Graphics Processing Unit), the compiler needs to check whether the size of the array is smaller than the size of the shared memory (\textit{array\_size <= shared\_mem\_size}). If this expression evaluates to true, then the compiler can place the array in shared memory which might accelerate the execution of the code.
Optimizing compilers usually have an API function that simplifies or proves expressions. Caviar can be integrated in such compilers simply by modifying that function to call the Caviar TRS instead of the original compiler TRS.

Caviar's main contribution is the introduction of three novel techniques to accelerate the use of e-graphs to prove expressions:

\begin{itemize}
    \item Iteration Level Check for Equivalence (ILC): The goal of this technique is to stop e-graph expansion as soon as an expression is proved (i.e., it evaluates to true or false). Thus, this technique eliminates the need to continue expanding the e-graph until it is saturated.
    
    \item Pulsing Caviar: This heuristic balances between exploration and exploitation by focusing on the most interesting paths after a fixed amount of time spent exploring all the possible ones.
    
    \item Non-Provable Patterns Detection (NPPD): A technique that enables Caviar to detect non-provable expressions. This is done by checking periodically if any of the equivalent forms of the initial expression matches one of the predefined non-provable patterns. If a non-provable pattern is detected, the equality saturation algorithm stops.
\end{itemize}

Pulsing E-graphs is the most important contribution of the paper. E-graphs tend to quickly consume all the memory on the machine. This means that the best extractable expression is limited by how much the e-graph was able to grow, lowering the ability of e-graph based systems to prove or simplify expressions. Pulsing allows e-graphs to find optimized expressions far beyond the usual memory limits, allowing more expressions to be optimized. Since we target proving whether an expression evaluates to true or false, the other two heuristics (ILC and NPPD) come into play. They do not increase the ability of e-graphs to prove expressions but rather accelerate the proof.

We evaluate our solution on expressions extracted from the Halide compiler \cite{ragan-kelley_halide_2013} while compiling a set of Halide programs. Halide is a state-of-the-art optimizing compiler that relies on a TRS to simplify and prove its expressions. It is designed mainly for the area of image processing. We start by evaluating the impact of each of the three contributions separately before combining them and comparing them to the original algorithm of e-graph expansion. We show that our proposed techniques, when used together, accelerate proving expressions. We also show that Caviar can prove expressions that a state-of-the-art TRS (Halide's TRS) fails to prove while only being 0.68x slower.

\section{E-Graphs and Equality Saturation}
\subsection{Equality Graphs}
Equality graphs (e-graphs) are a data structure, based on directed acyclic graphs (DAC), that stores a set of terms and an equivalence relation over them.
They are composed of E-Nodes (Equality Nodes). An E-Node is a function symbol (can be a constant or a term) that has e-classes as children. An E-Class (Equality Class) is a set of E-nodes, it represents the different equivalent forms that a term can take. An e-graph is a hierarchy of e-classes representing the different equivalent forms of the main term and the sub-terms included in it. 

\begin{figure}[hbt!]
    \centering
    \includegraphics[width=0.4\textwidth]{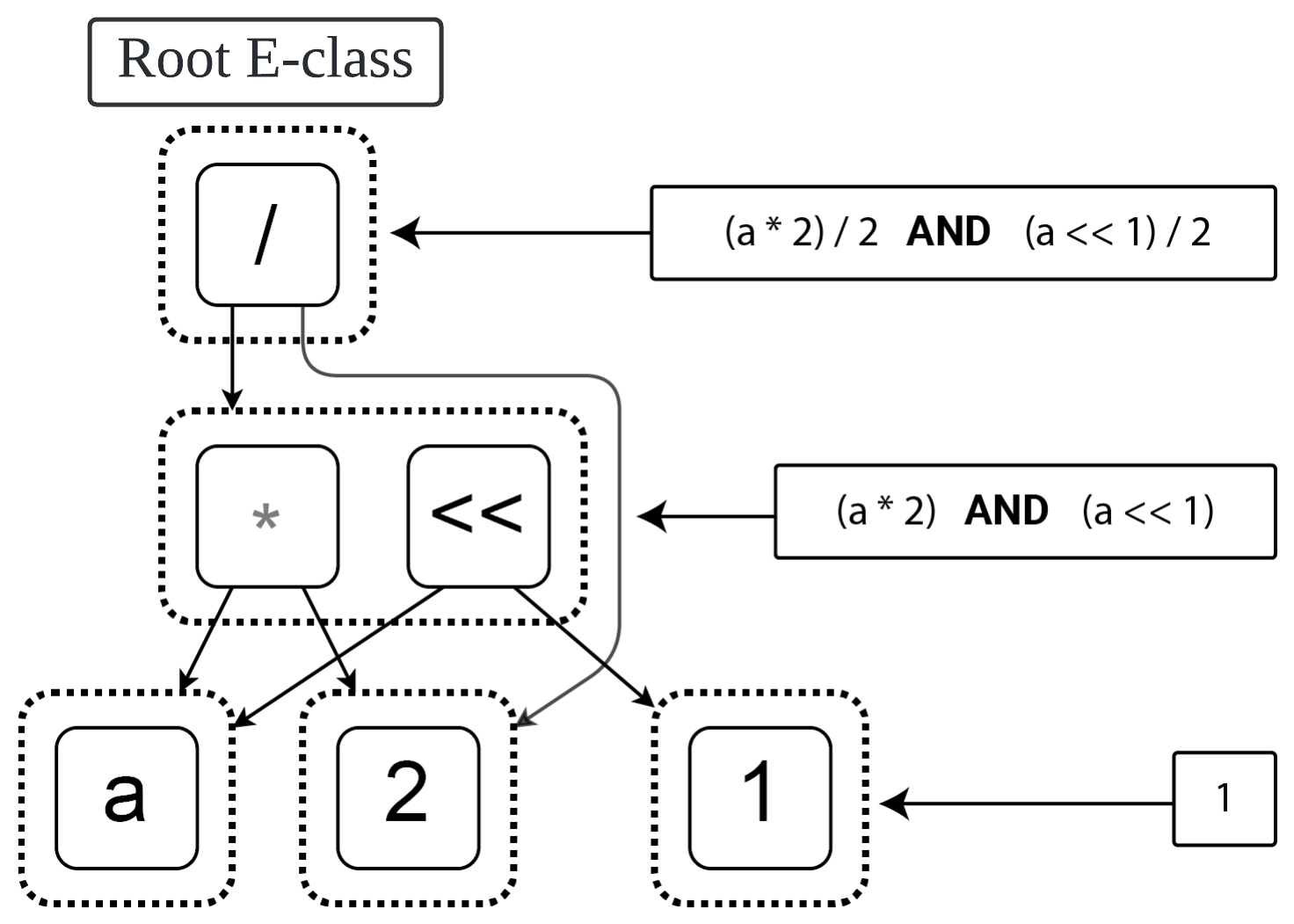}
    \caption{Example of an E-Graph containing the equivalent forms of $(a * 2) / 2$ \cite{willsey_egg_2021}}
    \label{fig:egraphExample}
\end{figure}

\autoref{fig:egraphExample} shows an example of an e-graph that contains the representations of the equivalent terms of $(a * 2 ) / 2)$. The nodes bordered with a solid line are E-Nodes while the nodes bordered with a dashed line are E-Classes.
The root e-class of the e-graph contains a single e-node that represents two equivalent forms of the initial expression: $(a * 2)/2$, and $(a \ll 1)/2$.

\subsection{Equality Saturation}

\begin{figure*}[ht]
  \centering
  \includegraphics[width=1\textwidth]{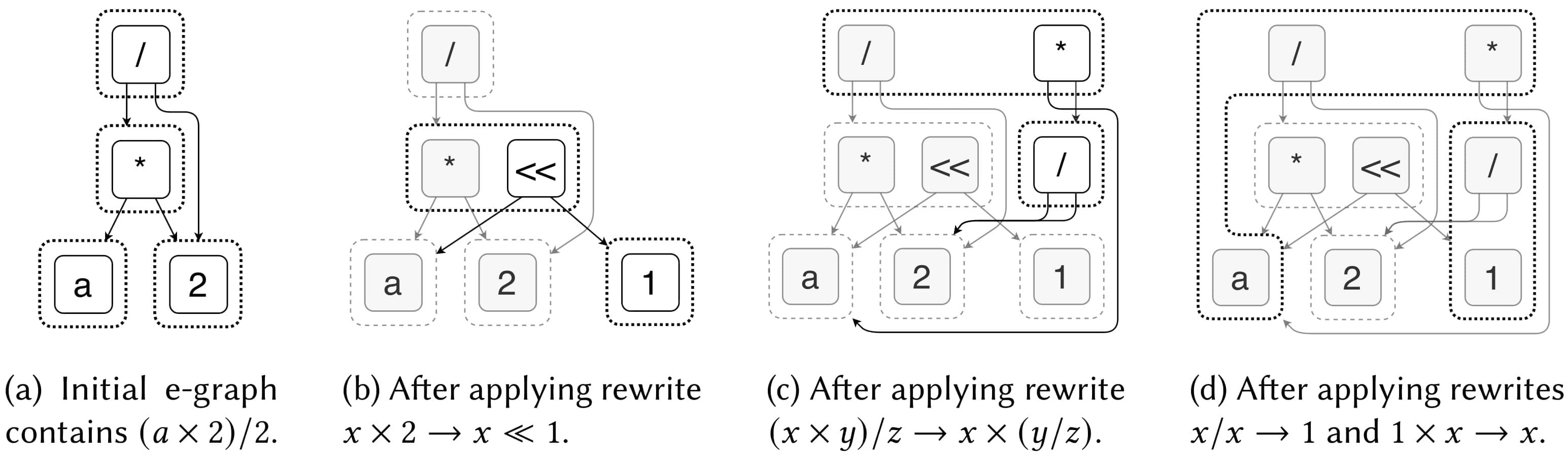}
  \caption{Example of term rewriting \cite{willsey_egg_2021}}
  \label{fig:egg_example}
\end{figure*}

Equality Saturation (ES) is a technique based on equality analysis that simply adds equality information to a common intermediate representation (IR)\footnote{In our case the intermediate representation is the e-graph.} without losing the original expression \cite{tate_equality_2009}. In our context, ES can be defined as applying all the rewriting rules simultaneously again and again until they no longer add information to the e-graph (i.e., until the e-graph no longer changes). We call this the saturation of an e-graph. The workflow of equality saturation is as follows: (1) Initialize an e-graph from the initial term; (2) Keep applying the set of rewriting rules until the e-graph is saturated (does not change anymore) or a timeout is reached.

\autoref{fig:egg_example}
shows the application of several rewrite rules to the term ($(a * 2)/2$). Rewriting in e-graph based TRSs is only additive and not destructive (unlike TRSs that are not based on e-graphs). The initial state of the term is kept in the e-graph even after the rewrite rules are applied. Thus applying a rule can only add information to the e-graph while conserving all the previous states. Applying the rewrite rule $x * 2 \implies x << 1 $ (the transition from \autoref{fig:egg_example}(a) to \autoref{fig:egg_example}(b)) for example, only added the e-node of the shift operation and that of the term 1. It is worth noting that all the previous nodes from the initial e-graph are conserved. We apply more rules in \autoref{fig:egg_example}(c) and \autoref{fig:egg_example}(d). The state represented in \autoref{fig:egg_example}(d) represents the saturated e-graph using the four used rules, none of the rules can be applied another time.

\subsection{E-graphs Based Term Rewriting Systems}
The non-destructive rewrites of e-graphs, if well used, can help implement an e-graph based TRS that has more potential than traditional ones:
\begin{itemize}
    \item Rewrites in e-graph based TRSs are unordered. This is unlike traditional rewriting algorithms where "later rewrites in the given rewrite list are favored in the sense that
    they can see the results of earlier rewrites" \cite{willsey_egg_2021}. This means that in traditional rewriting algorithms the result depends on the order of the rewrite list.
    \item After saturation, the E-Graph contains every possible derivation of the input term from the set of rules used to saturate the e-graph.
    \item Unordered rewrites combined with ES enable the composition of complex rules based on simple axiomatic rules thus reducing the size of the ruleset (rewriting rules).
\end{itemize}

\noindent Though highly useful, e-graph based TRSs are not without limitation:
\begin{itemize}
    \item The equality saturation algorithm is \textbf{computationally expensive} compared to greedy algorithms usually implemented in traditional TRSs.
    
    \item \textbf{The size of the e-graph can explode} if the ruleset contains non-axiomatic rules.
    
    \item Extracting the most optimized form of the input expression from the e-graph may be computationally expensive especially if the e-graph is large.
\end{itemize}

\section{Building Caviar}
\label{ch:caviar}

In this section, we describe the vanilla Caviar (Caviar without our three novel techniques that accelerate e-graph expansion). We built Caviar on top of the egg library \cite{willsey_egg_2021}, an open-source library that provides a state-of-the-art implementation of e-graphs and the equality saturation algorithm.
First, we show how we used egg to build Caviar, then we show how we built our ruleset.


\subsection{Vanilla Caviar}
The original equality saturation algorithm keeps on applying all the rewrite rules until the e-graph is saturated or a timeout is reached. Then, an extraction method is run to extract the best possible representation of the input expression based on a given metric. By default, the egg library uses the depth of the expression's AST as a metric.

To use Caviar for proving expressions, all that is needed is to switch the extraction method with another one that checks the equivalence between the root e-class and true or false.
It will check whether true or false matches one of the representations of the root e-class of the resulting e-graph. Algorithm \ref{fig:eqsatprv} shows the modified equality saturation algorithm. We mainly replaced the extraction method that was on line 9 with another method that checks for equivalence: \texttt{check\_equivalence()}. We use the same equivalence check provided by the egg library. It works by checking if the expression passed as a parameter is included in the root e-class of the e-graph of the expanded expression. For our implementation, we edited the function to return, on top of the data returned in the original implementation, a boolean indicating whether one of the goals matched, and the index of the matched goal.

The overhead of \texttt{check\_equivalence()} should be minimal so that it can be executed multiple times without penalizing the overall execution time.

\newsavebox{\codeboxc}
\begin{lrbox}{\codeboxc}
\begin{lstlisting}[language=egg,basicstyle={\ttfamily\footnotesize},escapechar=@]
function equality_saturation(expr, rewrites):
    egraph = initial_egraph(expr)
    while not egraph.is_saturated_or_timeout():
        // Apply the rewrite rules to the e-graph
        for rw in rewrites:
            egraph.apply(rw)
            
    // Check if true or false match the root class of egraph
    return egraph.check_equivalence([false, true])
\end{lstlisting}
\end{lrbox}

\begin{algorithm}[ht]
\SetAlgoLined
        \usebox{\codeboxc}
    \caption{Equality saturation algorithm utilized to prove expressions}
    \label{fig:eqsatprv}
\end{algorithm}

Algorithm \ref{fig:eqsatprv} proves an expression by expanding the initial e-graph of that expression through equality saturation, then it checks whether \texttt{true} or \texttt{false} are included in the root e-class of the resulting e-graph.

\subsection{Developing Caviar's Ruleset}

Building an optimal ruleset is a challenging task. On one hand, adding a large number of rules can slow down the TRS, and a considerable amount of those rules are unlikely to be used. On the other hand, restricting the ruleset to a small number of rules can limit the ability of the TRS to prove expressions.

We decided to only include axiomatic rules backed by the idea that the equality saturation algorithm will be able to progressively combine them while iterating thus deriving compound rules to prove complex patterns of expressions.
Our ruleset is composed of 132 axiomatic rules making it 7x smaller than Halide's one \footnote{We only consider rules that contain algebraic or boolean operators and have no Halide specific functions.}.

\section{Proposed Techniques to Improve Expression Proving}
\label{contributions}

In this section we describe three techniques we developed to improve the performance of Caviar in both the number of expressions it can prove and the execution time it takes: 
\begin{itemize}
    \item Iteration Level Check (ILC) for Equivalence: the goal of this technique is to stop e-graph expansion as soon as an expression is proved (i.e., it is found to be equivalent to true or false).
    
    \item  Pulsing Caviar: a heuristic that enhances both the number of proved expressions and the execution time of Caviar. 
    \item Non-Provable Patterns Detection (NPPD): a technique that enables Caviar to detect expressions that cannot be proven.
\end{itemize}

\subsection{Iteration Level Check for Equivalence}
\label{ILC}
The main idea of ILC (Iteration Level Check for Equivalence) is as follows: 
at some iteration \emph{t} of the equality saturation algorithm, the e-graph reaches a state where Caviar could prove the input expression but it continues execution since it did not reach saturation. The execution-only stops when a stopping mechanism is reached (saturation is reached or a timeout).

In fact, in most cases for expressions that Caviar can prove, it can prove those expressions long before reaching saturation. In addition, most of the execution time is spent trying to either reach saturation or a stopping mechanism, and that time is wasted time.

To avoid this, ILC stops the equality saturation algorithm as soon as the input expression is proven (i.e., it is found to be equivalent to either true or false).

Algorithm \autoref{fig:iterFlowChart} illustrates how we modified the saturation algorithm to implement ILC. We mainly check for the equivalence\footnote{The time for equivalence check is in the order of nanoseconds which is negligible} at the outer loop of equality saturation, just after applying all the rewrites.

\begin{lrbox}{\codeboxc}
\begin{lstlisting}[language=egg,basicstyle={\ttfamily\footnotesize},escapechar=@]
function equality_saturation(expr, rewrites):
    egraph = initial_egraph(expr)
    while not egraph.is_saturated_or_timeout():
        for rw in rewrites:
            egraph.apply(rw)
        
        // Check if true or false match the root class of  the egraph
        result = egraph.check_equivalence([false, true])
        if (result.matched)
            return result

    // If we reach saturation or time out do one last check
    return egraph.check_equivalence([false, true])
\end{lstlisting}
\end{lrbox}

\begin{algorithm}[ht]
\SetAlgoLined
    \usebox{\codeboxc}
    \caption{Modified equality saturation algorithm to include iteration level check for true or false}
    \label{fig:iterFlowChart}
\end{algorithm}

\subsection{Pulsing Caviar}
We successfully accelerated proving expressions in Caviar through ILC. But ILC only accelerated expressions Caviar was capable of proving. For expressions that were either non-provable or Caviar lacked the power to prove, Caviar rarely reached saturation and almost always failed due to the time limit set making Caviar's performance bounded by the time limit set.

Caviar uses equality saturation which favors exploration heavily. In each iteration, Caviar matches all the possible rewrites then apply them. This ensures that it explores all the possible paths simultaneously thus rendering the order of the rewrites obsolete. Caviar exploits the different paths throughout the iterations. Each iteration can be considered as a step in exploring the search space.

From our first experiences, we noticed that Caviar can extract a much shorter equivalent expression after a small amount of time. But since equality saturation favors exploration, the paths leading to these short expressions would only be exploited after several iterations. 

The main idea of Pulsing Caviar is to choose the most promising paths (based on the size of the expression) and focus on them rather than all the e-graph. Caviar prunes the e-graph after a fixed amount of time (threshold) leading to more exploitation in these paths and potentially reducing the execution time for expressions we can prove and those we cannot.

To implement this technique, we repeatedly (i) stop the equality saturation algorithm once we reach a threshold (that we define empirically), (ii) 
extract the expression that has the smallest number of AST nodes
(using the number of AST nodes is both fast and proved to provide good results), (iii) reinitialize the e-graph with the newly extracted expression, then (iv) re-run the equality saturation algorithm another time on the newly created e-graph. We keep repeating the previous steps until Caviar reaches the time limit set. At most Caviar will run the equality saturation $(time\_limit/threshold)$ times.

Each time the e-graph is reinitialized, the best expression progresses "in pulses" towards shorter expressions while better exploiting the execution time when an ordinary e-graph would have run out of memory and would have tried to expand unused parts of the e-graph. Algorithm \ref{fig:beh_flowchar} illustrates the different steps of this approach.

By choosing to focus on the more promising paths in the e-graph, Pulsing Caviar might prevent Caviar from proving an expression. This situation happens if one of the pulses prunes the path leading to the solution. The results from our experiments show that this is rarely the case. In most cases, Caviar manages to prove the expressions in a shorter time, which explains the speedup gained and the increase in the number of expressions proved. 

\begin{lrbox}{\codeboxc}
\begin{lstlisting}[language=egg,basicstyle={\ttfamily\footnotesize},escapechar=@]
function equality_saturation(expr, rewrites, threshold):
    egraph = initial_egraph(expr)
    while not egraph.is_saturated_or_timeout():
        for rw in rewrites:
            egraph.apply(rw)

        // If the threshold is reached, reinitialize egraph
        // with the shortest extracted expression
        if (check_threshold(threshold))
            egraph = egraph.best_expression()
            
    // If we reach saturation or time out do one last check
    return egraph.check_equivalence([false, true])
\end{lstlisting}
\end{lrbox}

\begin{algorithm}[ht]
\SetAlgoLined
        \usebox{\codeboxc}
    \caption{Modified equality saturation algorithm to include Pulsing Caviar}
    \label{fig:beh_flowchar}
\end{algorithm}

\subsection{Non-Provable Patterns Detection}
Despite the ability of e-graphs to represent equivalence relations efficiently, executing the equality saturation algorithm with our ruleset rarely reaches saturation. This happens mainly since some expressions have an infinite number of equivalent forms that can be derived using the rewrite rules we defined. Some of these expressions are non-provable (i.e., Caviar cannot decide about whether they are true or false). An example of such expressions is whether $x \neq c$ (where $x$ is a variable and $c$ is a constant). Caviar, by default, does not have any information about the possible values of $x$ and therefore cannot decide whether $x \neq c$.

The main idea of the proposed technique is to detect non-provable expressions and stop the equality saturation algorithm early when a non-provable expression is detected. This technique is only used while proving expressions and is not used if Caviar is used for simplification purposes.
When defining the set of non-provable patterns, we only include the most frequent non-provable patterns of expressions that Caviar faces when proving expressions. We modified the equality saturation algorithm to check for these patterns in the e-graph after each iteration and stop if one of them is found.

Algorithm \ref{fig:patterns_flowchart} shows the details of the proposed technique.

\begin{lrbox}{\codeboxc}
\begin{lstlisting}[language=egg,basicstyle={\ttfamily\footnotesize},escapechar=@]
function equality_saturation(expr, rewrites, patterns):
    egraph = initial_egraph(expr)
    while not egraph.is_saturated_or_timeout():
        for rw in rewrites:
            egraph.apply(rw)

        // Check if a pattern matches. If so, return
        // the best expression that can be extracted
        result = egraph.check_equivalence(patterns)
        if (result.matched)
            return egraph.best_expression()
            
    // If we reach saturation or time out do one last check
    return egraph.check_equivalence([false, true])
\end{lstlisting}
\end{lrbox}

\begin{algorithm}[ht]
\SetAlgoLined
        \usebox{\codeboxc}
    \caption{Modified equality saturation algorithm to include the NPPD technique}
    \label{fig:patterns_flowchart}
\end{algorithm}

For each non-provable pattern, we define the conditions that make it non-provable. For example, expressions of the form $c < a\;\%\;b$ are non-provable only if $c < |b|$, otherwise, they can be proven. The expression $8 < x \; \% \; 8$ matches the pattern but it is always false since it doesn't satisfy the condition.

\subsection{Hyper-parameter Tuning}
\label{tuning}

Caviar is flexible by design and can be parametrized through multiple hyper-parameters. In this section, we describe these hyper-parameters and show how we tuned their values.
To tune the hyper-parameters of Caviar, we generated automatically a set of expressions on which we ran our experiments. We call this dataset the \textit{tuning dataset}. These expressions were generated by the Halide compiler while compiling random programs. These random programs were generated using the Halide random code generator \cite{adams_learning_2019}. This is a random code generator that was developed to train a deep learning cost model for the Halide auto-scheduler and was designed to generate programs that are similar to realistic programs in the areas of image processing and deep learning.

\begin{itemize}
    \item \textbf{Choosing a Time Limit for Equality Saturation.} In order to tune the time limit of the equality saturation algorithm, we run the following experiment. We run Caviar on the tuning dataset with different values for the time limit ranging from 0.001s up to 60s. The goal was to explore the impact of the time limit on the performance of Caviar (its ability to simplify expression and its overall execution time). Our results show that the best values for the time limit depend on the goal of the user. If speed is important, then a time limit of 1s is preferred. To improve the ability of Caviar to simplify expressions, time limits ranging from 3s to 10s are preferred depending on how much execution time is allowed.
    
    \item \textbf{Choosing a Threshold for Pulsing Caviar.} We experimented with different values for the threshold: 0.01s, 0.05s, 0.1, 0.25s, 0.5s, 0.75s and 1s.  The results show that the best values were 0.01s and 0.05s, with 0.05s being the best compromise between the number of expressions Caviar can prove and the speedup achieved.

    \item \textbf{Choosing the List of Non-provable Patterns.} Our goal was to identify the list of non-provable patterns to use in the NPPD method. We first created a list of 17 non-provable patterns (we found these patterns by inspecting the expressions that Caviar could not simplify). We then noticed that only 5 patterns were used frequently and thus limited our list of patterns to these 5.
\end{itemize}

\section{Evaluation}
All experiments were run on a cluster of 24 nodes. Each node has an Intel Xeon E5-2695 v2 @ 2.40GHz (two sockets with 12 cores per socket). Each node has 128GB of memory.

In order to evaluate our TRS, we compare it to the TRS of the Halide~\cite{ragan-kelley_halide_2013} compiler (a state-of-the-art TRS). Through its API, Caviar can be integrated into different systems that need to simplify or prove expressions. In the case of the Halide compiler, Halide calls a specific function for proving expressions called \texttt{can\_prove}, and for simplifying expressions it calls \texttt{simplify}. To perform our evaluation, we modified the Halide call to prove expressions (\texttt{can\_prove}) so that it uses the Caviar TRS instead of the original Halide TRS.

\subsection{Data Sets}

\paragraph{Test Dataset} This dataset contains 5000 expressions generated automatically. These expressions were generated by the Halide compiler using the same method described in Section\ref{tuning} (they are different from the expressions of the tuning dataset though).
The goal of using this dataset is to evaluate Caviar as well as the three proposed techniques. It is also used to evaluate whether the hyper-parameters we chose in Section \ref{tuning} generalize. The test dataset is used in all of the experiments of the evaluation section except the one in Section \ref{hard}.

\paragraph{Hard Dataset} We also wanted to evaluate Caviar's performance on complicated expressions, expressions that Halide fails to prove. In the test dataset, we only had 949 expressions that Halide failed to prove. Since this number is relatively small, we created a new dataset that has a higher number of expressions that Halide fails to prove. We compiled this new dataset using the same method that we used to generate the Test dataset. We then filtered out the expressions that Halide was capable of proving, resulting in a dataset of 5787 expressions that Halide could not prove. This dataset is only used in Section \ref{hard}.

\subsection{Evaluating Caviar\label{sechyper}}
We first evaluate the vanilla Caviar (the work described in \autoref{ch:caviar} and which includes the basic Caviar without our three proposed techniques).

Since most of our efforts in this work were dedicated to optimizing the time taken by our TRS, we analyze the impact of changing the time limit imposed on the equality saturation algorithm in Caviar. For the other two limits: iterations limit and e-nodes limit, they are set to a very high limit.
The goal of this experiment is to evaluate different time limits and pick one that will be used in the next experiments as a time limit.

\begin{table}[ht]
    \caption{Number of proved expressions by vanilla Caviar over different time limits}
    \label{fig:vanilla_exprs}
    \begin{center}
        \begin{tabular}{ @{}c | c@{}  }
            \toprule
            \textbf{Time Limit (s)} & \textbf{\# of expressions proved \;}\\
            \toprule
            1   &   4427    \\
            \midrule
            2   &   4433    \\
            \midrule
            3   &   4443    \\
            \midrule
            5   &   4447    \\
            \bottomrule
        \end{tabular}
    \end{center}
\end{table}

\autoref{fig:vanilla_exprs} shows the results of the number of expressions proved in each experiment each with a different time limit. The number of proved expressions improves slightly when the time limit goes up (as one would expect). This shows that some expressions can require up to 5 seconds or even more to be proven.

For the rest of the evaluations, we set the time limit to 3s to balance the execution time and the number of proved expressions.

\subsection{Caviar with ILC}
The first contribution we are going to evaluate is ILC (Iteration Level Check for Equivalence). ILC, as described in \autoref{ILC}, reduces the execution time for expressions that can be proven by our TRS. It does that by stopping the execution once a goal is found in the e-graph.

\autoref{tab:evalilc} shows the time taken by Caviar with and without ILC to prove all expressions of our test set. It also shows the total time spent on expressions Caviar can prove, and the total number of proved expressions out of the test set.
\begin{table}[ht]
    \caption{Evaluation of ILC}
    \label{tab:evalilc}
    \begin{center}
        \begin{tabular}{ @{}l | c@{} | c@{} }
            \toprule
            \textbf{} & \textbf{With ILC \;} & \textbf{Vanilla \;} \\
            \toprule
            Time for all expressions (s)  & 1049.200	 &   14818.459       \\
            \midrule
            Time for proved expressions (s) &  49.854 & 13821.738    \\
            \midrule
            Number of proved expressions &   4443 & 4443      \\
            \bottomrule
        \end{tabular}
    \end{center}
\end{table}

This experiment shows that a \textbf{277x} speedup is gained on expressions that could be proven. And since they represent about \textbf{88\%} of the dataset, a \textbf{14x} speedup is gained on the total execution time. The number of proved expressions remains the same because this method only affects the execution time for expressions that Caviar can prove.

\subsection{Pulsing Caviar}
\label{beh_results_sec}
In the next experiment, we evaluate the pulsing Caviar technique on our test set. 

\begin{figure}[ht]
     \centering
     \begin{subfigure}[b]{0.5\textwidth}
        \centering
            \includegraphics[width=1\textwidth]{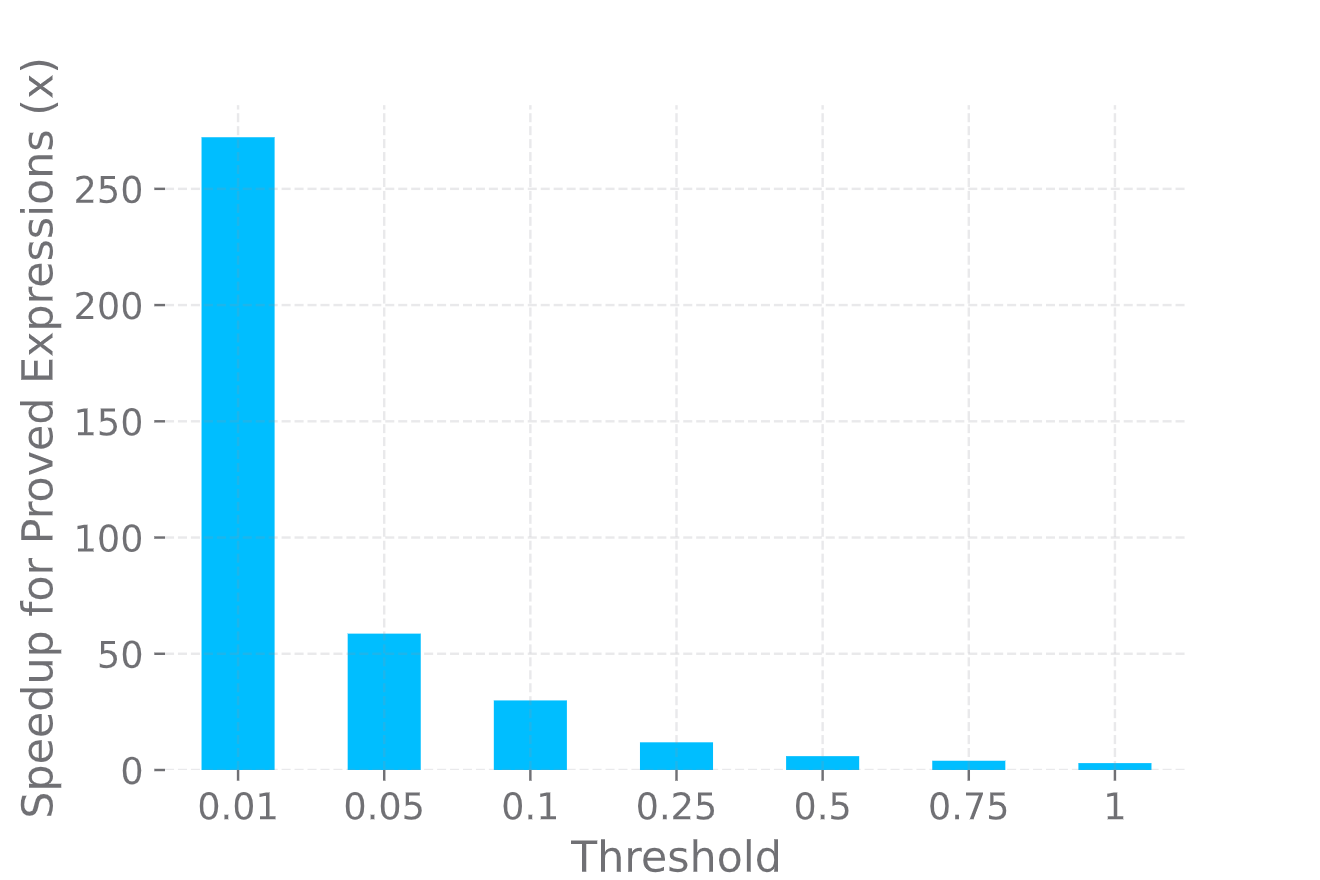}
            \caption{Speedup of proved expressions for different threshold.}
     \end{subfigure}
     \hfill
     \begin{subfigure}[b]{0.5\textwidth}
        \centering
            \includegraphics[width=1\textwidth]{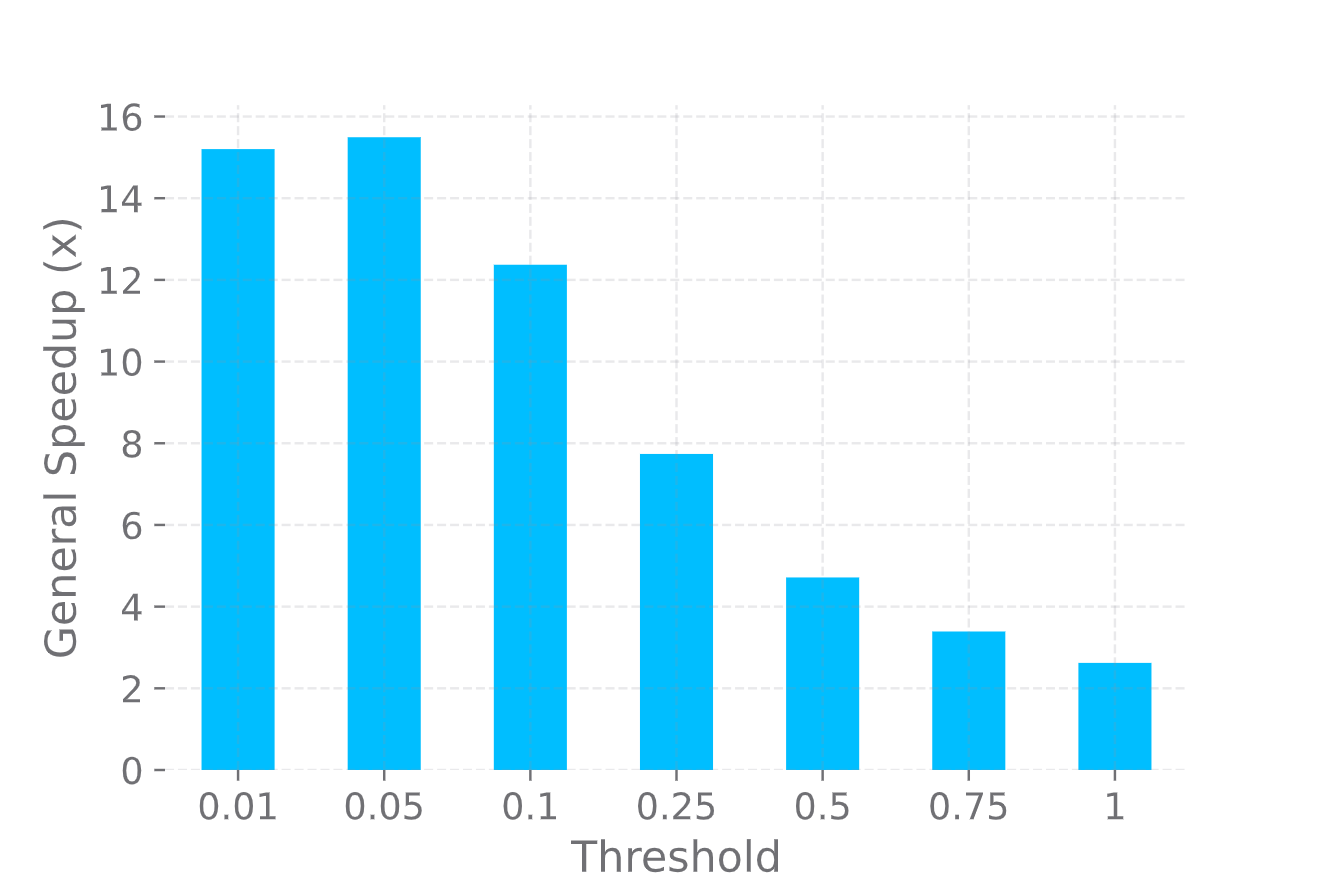}
            \caption{General speedup for different values of the threshold}
     \end{subfigure}
        \caption{Speedups for different values of the threshold on the test set.}
    \label{fig:beh_eval_exectimes}
\end{figure}

We mainly evaluate the effect of different thresholds (pulsing Caviar thresholds) on the speedup of proving expressions. The results are shown in \autoref{fig:beh_eval_exectimes}. \autoref{fig:beh_eval_nbrProved} shows the number of expressions that Pulsing Caviar can prove for each value of the threshold.
The main takeaways from these experiments are as follows:

\begin{itemize}
    \item The speedup on expressions we can prove is negatively correlated with the value of the threshold, as illustrated in  \autoref{fig:beh_eval_exectimes}(a). We can see that for the threshold of \textbf{0.01s}, Pulsing Caviar is \textbf{272.2x} faster than the vanilla Caviar, yet it still proves the same number of expressions that vanilla Caviar proves as shown in \autoref{fig:beh_eval_nbrProved}.
    
    \item The general speedup is also negatively correlated with the different values of the threshold. This is illustrated in \autoref{fig:beh_eval_exectimes}(b). It is worth noting that for all the values of the threshold, Pulsing Caviar is at least \textbf{2.62x} faster than the vanilla Caviar. The fastest one are \textbf{0.01s} and \textbf{0.05s}, as they achieve a \textbf{15.49x} speedup while improving the number of expressions they can prove by \textbf{57}.
    
    \item The number of expressions proved by the pulsing Caviar heuristic are all better than the baseline implementation. The results are illustrated in \autoref{fig:beh_eval_nbrProved}. The \textbf{0.25s}, \textbf{0.5s}, and \textbf{0.75s} thresholds all give the best number of proved expressions (around \textbf{4512} expressions proved), that is around \textbf{69} expressions more than the vanilla Caviar.
\end{itemize}

\begin{table}[ht]
    \caption{Number of proved expressions over different values of the threshold on the test set}
    \label{fig:beh_eval_nbrProved}
    \begin{center}
        \begin{tabular}{ @{}l | c@{} }
            \toprule
            \textbf{Threshold (s)} & \textbf{\# proved expressions \;}  \\
            \toprule
            
            0.01 &   4443       \\
            \midrule
            0.05 &  4500    \\
            \midrule
            0.1 &  4498    \\
            \midrule
            0.25 &  4511    \\
            \midrule
            0.5 &  4512    \\
            \midrule
            0.75 &  4513    \\
            \midrule
            1 &  4508    \\
            \midrule
            Vanilla &  4443    \\
            \bottomrule
        \end{tabular}
    \end{center}
\end{table}

All of the thresholds manage to score significant improvements in both the execution time and the number of expressions proved.

\subsection{Caviar with Non-Provable Patterns Detection}

\autoref{fig:patterns_eval_time} shows the total execution time, the number of proved, non-provable, and not proved expressions for the two versions of Caviar: the vanilla Caviar and Caviar with NPPD.
 
 \begin{table}[ht]
   \caption{Total execution time with and without NPPD}
    \label{fig:patterns_eval_time}
    \begin{center}
        \begin{tabular}{ @{}l | c@{} | c@{} }
            \toprule
            \textbf{} & \textbf{Vanilla \;} & \textbf{With NPPD \;} \\
            \toprule
            Total time for all expressions (s)  & 14818.45	 &   14778.21       \\
            \midrule
            \#  proved expressions &   4443 & 4443      \\
            \midrule
            \#  expressions not proved &   557 & 346      \\
            \midrule
            \#  non-provable expressions &   - & 211      \\
            
            \bottomrule
        \end{tabular}
    \end{center}
\end{table}

We notice that the method, applied on its own, has approximately no effect on the execution time (a speedup of 1.002x), but it makes the TRS capable of identifying \textbf{38\%} of the non proved expressions as non-provable ones (as shown in \autoref{fig:patterns_eval_time}). From the results above, we can conclude what follows:
\begin{itemize}
    \item The non-provable patterns detection method does not improve the execution time on expressions issued from the Halide compiler, we only notice a 1.002x speedup, which can be explained by the small ratio of expressions that do match the non-provable patterns.
    \item The method allows Caviar to differentiate between expressions that the TRS cannot prove and the expressions that are known to be non-provable.
\end{itemize}

\subsection{Summary of the Results}
To summarize the results we could gather from all the experiments:
\begin{itemize}
    \item The ILC technique reduces considerably the execution time. Caviar with ILC is \textbf{14x} faster.
    \item The Pulsing Caviar technique helped in improving both the number of expressions Caviar can prove and the execution time. Pulsing Caviar leads to speedup of \textbf{15x} while also increasing the number of expressions Caviar can prove.
    \item The Non-Provable Pattern detection technique enables Caviar to detect expressions that cannot be proved without adding any additional execution time.
\end{itemize}

\subsection{Combining Contributions}
Our goal in this section is to evaluate the impact made once we add the three proposed techniques together in our TRS at the same time, as well as to evaluate the proposed TRS compared to a state-of-the-art industrial TRS (Halide's TRS).

\subsubsection{Comparing Caviar+ and Vanilla Caviar}
Caviar+ is the TRS resulting from using all the previously mentioned techniques together. Namely, ILC, Pulsing Caviar, and NPPD. 

We compare three variants of this optimized TRS: the $1^{st}$ is Caviar with ILC only, the $2^{nd}$ is Caviar+ with 0.25s as a threshold for Pulsing Caviar (this threshold gave the best number of proved expressions), and the $3^{rd}$ is Caviar+ with 0.1s as a threshold for Pulsing Caviar (this threshold gave the best speedup against Vanilla Caviar).

\begin{figure}[ht]
    \centering
    \includegraphics[width=0.5\textwidth]{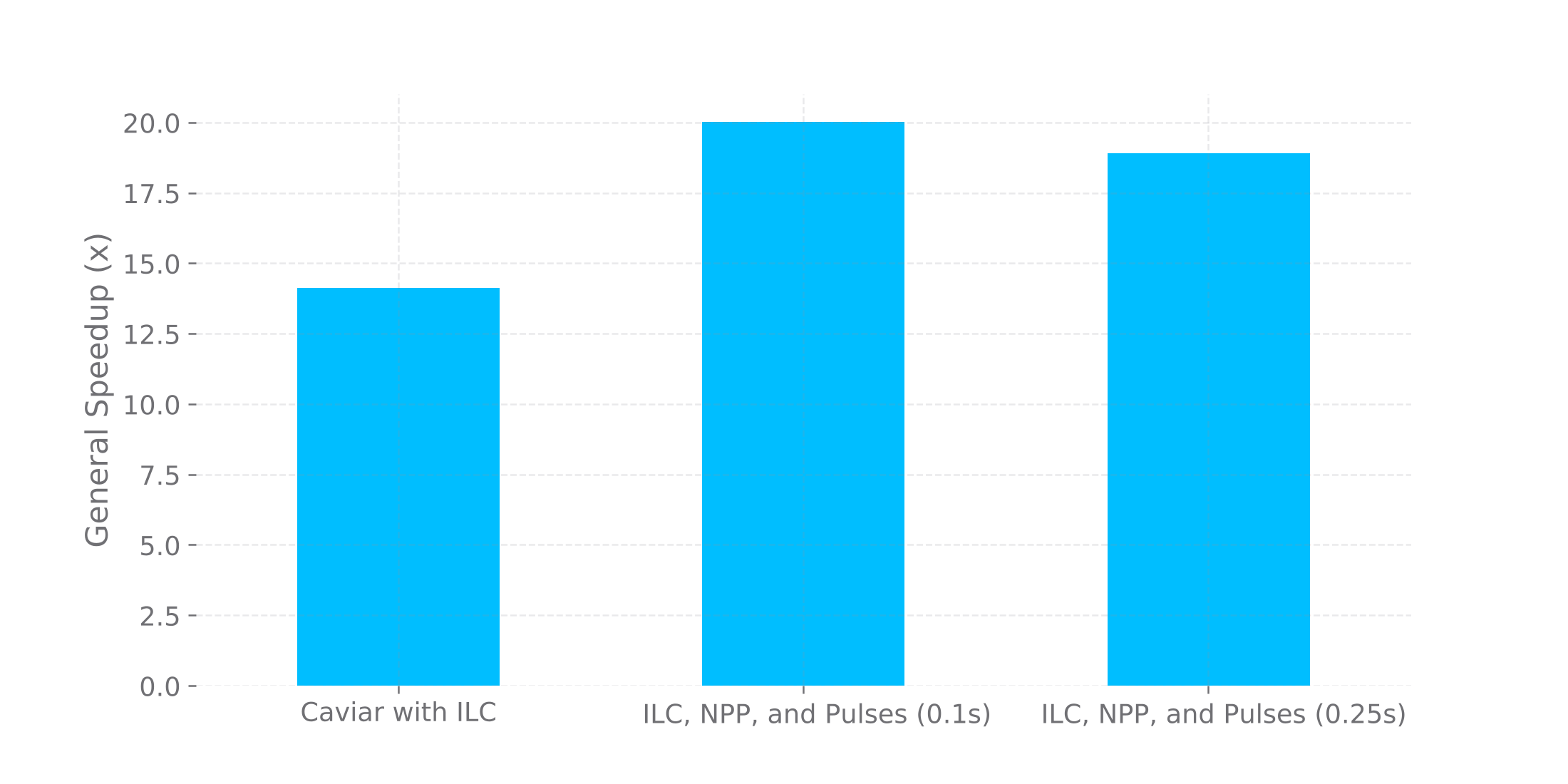}
    \caption{General Speedup for Caviar's variants}
    \label{fig:mov_eval_exectimes}
\end{figure}

\begin{figure}[ht]
    \centering
    \includegraphics[width=0.5\textwidth]{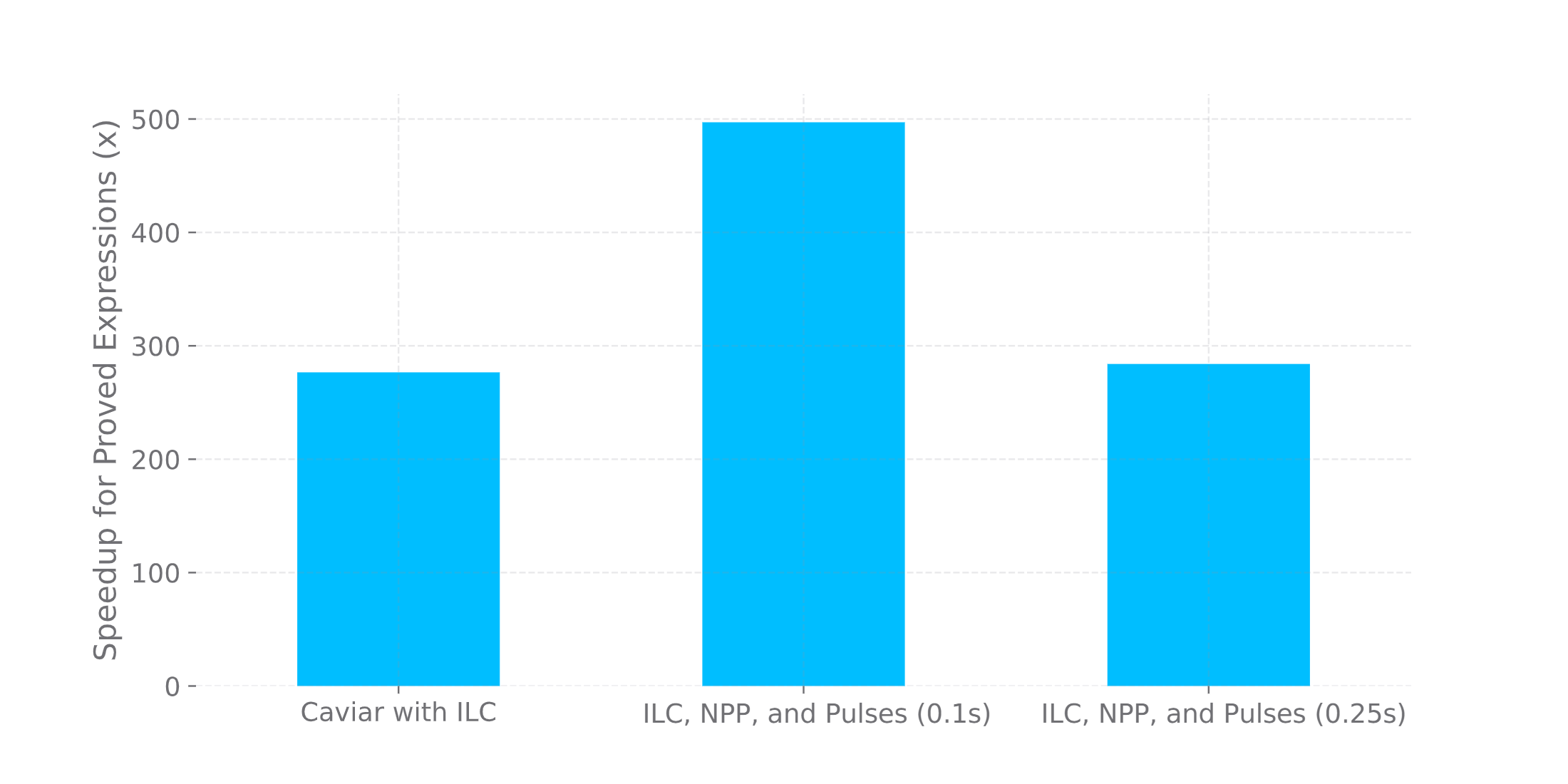}
    \caption{Speedup for proved expressions for Caviar's variants}
    \label{fig:mov_eval_exectimes_proved}
\end{figure}

\autoref{fig:mov_eval_exectimes} shows the speedups for each of these three versions compared to the Vanilla Caviar on all expressions. \autoref{fig:mov_eval_exectimes_proved} shows the speedups but on the proved expressions only, while \autoref{tab:caviar_variants} shows the number of expressions that be can be proven along with the execution times.

\begin{table}[ht]
    \caption{Auto-tuning the time-limit to compare with Halide}
    \label{tab:caviar_variants}
    \begin{center}
        \begin{tabular}{ @{}l | c@{} | c@{} | c@{} }
\toprule
{} &     Time for Proved (s) \, &    Time (s) \, &  \# Proved  \\
\toprule
Caviar (ILC) &      49.97 &     1048.04 &      4443 \\
\midrule
Caviar+ (0.1s)  &      27.80 &      740.22 &      4508 \\
\midrule
Caviar+(0.25s)  &      48.67 &      783.95 &      4510 \\
\midrule
Vanilla         &   13821.74 &    14818.46\, &      4443 \\
\midrule
Halide          &       0.52 &        1.16 &      4051 \\
\bottomrule
\end{tabular}
    \end{center}
\end{table}

The results show that we have gained a speedup of \textbf{20x} on all the expressions and \textbf{497x} on the proved expressions when comparing Vanilla Caviar with the fastest of the two proposed variants (the one that uses a Pulsing Caviar threshold of 0.1s). We also notice that the variant that uses a Pulsing Caviar threshold of 0.25s can prove \textbf{90\%} of the expressions, while the vanilla Caviar can only prove \textbf{88\%} of them.

This experiment shows that the combination of the three contributions (ILC, Pulsing Caviar, and NPPD) results in making Caviar+ significantly faster than the vanilla Caviar and able to prove more expressions as well. 

\subsection{Comparing Caviar+ and Halide's TRS\label{sechard}}
The last evaluation compares Halide's TRS and Caviar+. As we have seen before, the time limit imposed on Caviar affects its execution time and affects how many expressions it can prove. So before comparing Caviar+ with Halide, we are first going to determine which time limit to use in order to have a total execution time comparable to that of Halide. \autoref{tab:caviar_tl} shows the results of this experiment.

\begin{table}[ht]
    \caption{Identifying a time-limit appropriate for comparison with Halide}
    \label{tab:caviar_tl}
    \begin{center}
        \begin{tabular}{ @{}l | c@{} | c@{} | c@{} }
            \toprule
            \textbf{Timelimit\;} &	  \textbf{\# Proved\;}  &	    \textbf{Time (s) \,}     & 	    \textbf{Time for Proved (s) \,} \\
            \toprule
            0.001	  &   4015	    &       1.45           &    	0.42 \\
            \midrule
            0.002	  &   4054	    &       2.49           &    	0.62 \\
            \midrule
            0.003	  &   4128	    &       3.2            & 	    0.79 \\
            \midrule
            0.004	  &   4177	    &       3.64           &    	0.81 \\
            \midrule
            0.005	  &   4204	    &       4.27           &    	0.96 \\
            \midrule
            Halide    &   4052      &       1.16           &        0.51 \\
            \bottomrule
        \end{tabular}
    \end{center}
\end{table}

The experiments show that Caviar can be as fast as Halide if the time limit is set to 0.001s. For higher time limits, Caviar proves more expressions than Halide but it is slower.  We will use this time limit for our experiments.

We also decided to use the fastest Caviar variant namely Caviar+. \autoref{fig:caviar_v_halide} shows the results of the experiments, the main takeaway being:
\begin{itemize}
    \item Caviar alone proves fewer expressions than Halide.
    \item Caviar+ can prove more expressions than Halide and takes less time to do it (0.51s for Halide, 0.47s for Caviar).
    \item Caviar+ is still generally slower on expressions it cannot prove, it takes Caviar 1.7s to finish proving the test set while Halide takes 1.16s.
\end{itemize}

\begin{table}[ht]
    \caption{Comparison between Caviar and Halide based on the execution times and the number of expressions proved.}
    \label{fig:caviar_v_halide}
    \begin{center}
        \begin{tabular}{ @{}l | c@{} | c@{} | c@{} }
            \toprule
            \textbf{} & \textbf{Caviar \;} & \textbf{Caviar+ \;} & \textbf{Halide} \\
            \toprule
            Exec. Time (s)  & 1.45	 &   1.7    &   1.16       \\
            \midrule
            Exec. Time Proved &   0.42 & 0.47   &   0.51      \\
            \midrule
            \#  expressions proved &   4015 & 4061  &   4052      \\
            
            \bottomrule
        \end{tabular}
    \end{center}
\end{table}

\subsubsection{Evaluating Caviar on the Hard Dataset\label{hard}}

We also wanted to evaluate Caviar's performance on complicated expressions, expressions that Halide fails to prove. This is why we reevaluated\footnote{The time limit used in this evaluation is 3s} Caviar on the \textit{Hard Dataset}. Caviar was able to prove \textbf{2942 expressions}, which represents \textbf{51\%} of this dataset. These results show that Caviar can prove more expressions than Halide if given enough time.

\section{Related Work}
\textit{E-graphs}. Since their first appearance
\cite{nelson_fast_1980}, E-graphs have been used in different domains, although mainly focused on program optimization. E-graphs were initially proposed as a data structure that is capable of representing equivalence relation efficiently, before becoming an essential part of every SMT solver \cite{willsey_egg_2021}.

\textit{Equality saturation}. Equality Saturation is a technique based on equality analysis that simply adds equality information to a common intermediate representation (IR) without losing the original expression. Thus, after each equality analysis run, both the old expression and the new one are represented, each run of the equality analysis is the application of all rewrite rules \cite{tate_equality_2009}.

\textit{egg}.
Egg provides a robust open-source implementation of e-graphs and equality saturation. It is being used in multiple projects including 
Ruler \cite{DBLP:journals/corr/abs-2108-10436} which uses equality saturation to automatically infer rewrite rules, 
Diospyros \cite{10.1145/3445814.3446707} which performs vectorization for digital signal processors via equality saturation,
Tensat \cite{DBLP:journals/corr/abs-2101-01332} which performs tensor graph superoptimization using equality saturation and e-graphs, 
Herbie \cite{10.1145/2813885.2737959} which improves the accuracy of floating-point expressions,
Szalinski \cite{10.1145/3385412.3386012} which is a tool that uses equality saturation with semantics-preserving Computer-Aided Design (CAD) rewrites to efficiently search for more optimized equivalent programs, 
SPORES \cite{wang_spores_2020} which uses equality saturation to simplify linear algebra expressions, 
and Glenside \cite{DBLP:journals/corr/abs-2105-09377} which performs term rewriting using equality saturation to map program
fragments to hardware accelerator invocations and automatically discover classic data layout transformations.

\textit{Peggy Compiler}. Peggy \cite{stepp_equality_2011} is a compiler based on the equality saturation technique. It performs program optimizations as well as translation validation between program pairs. In order to optimize a program, Peggy, first transforms the program into an internal representation (IR) form called Program Expression Graphs (PEG); then it applies equality analysis on it. The set of equivalent programs is explored and the best program is picked (according to a predefined metric).

\textit{SPORES}. While Caviar uses equality saturation to simplify arithmetic expressions, SPORES \cite{wang_spores_2020} uses it to simplify linear algebra  expressions. SPORES first converts linear algebra expressions to relational algebra. The SPORES optimizer then uses equality saturation to explore the complete representation of the search space. The last step is the extraction of the optimal expression using a constraint-based solver.

\textit{Halide TRS} The Halide compiler relies internally on a term rewriting system to prove certain proprieties of code. In order to achieve that, it uses a custom algorithm. This algorithm represents an input expression as a directed acyclic graph (DAG). It simplifies this expression in a depth-first, bottom-up traversal of the expression's DAG. At each node, it attempts to match the expression with the LHS of the rules in a fixed order. When a match is found, the algorithm rewrites the subtree expression using the RHS of the matched rule, then it attempts to simplify the subtree expression again. If no rule matches the subtree, the traversal continues; when the entire expression cannot be simplified further, the rewritten expression is returned \cite{newcomb_verifying_2020}. This simplification algorithm is very fast since it's a greedy algorithm that has no backtracking, and it requires very little memory since it keeps only one expression in the state. But unlike our work, the order of applying rewrite rules is fixed to avoid cycles and guarantee termination. This is why axiomatic rules are avoided and are replaced with several variations of specific ones, driving the number of rules to increase significantly. The order of applying rewrite rules also makes the maintenance of the TRS harder, since adding rules at the wrong place may break the termination guarantee, contrary to our work where the order of rules doesn't matter.

\section{Conclusion}

In this work, we presented Caviar, an e-graph-based term rewriting system for simplifying and proving algebraic expressions. Caviar uses a ruleset that is composed of axiomatic rules. We introduced three techniques to enhance Caviar's performance: the Iteration Level Check for equivalence, and Pulsing Caviar reduced the execution times by a factor of more than 15x. While the non-provable patterns detection technique enhanced Caviar's ability to stop the execution once we are sure the expression cannot be proven instead of waiting to reach saturation or one of a time limit.
The proposed techniques allow Caviar to accelerate e-graph expansion by 20x for the task of proving expressions. They also allow Caviar to prove 51\% of the expressions that Halide’s TRS cannot prove while being only 0.68x slower.
In addition, Caviar is much more flexible in terms of the number of expressions it can prove and its execution time. By changing the time limit we can tune its performance and adapt it to different situations.
\balance
\bibliographystyle{ACM-Reference-Format}
\bibliography{sample-base}

\end{document}